\documentclass[aps,prl,epsfig,floats,twocolumn,superscriptaddress,amssymb,amsmath,floatfix]{revtex4}
\usepackage{amsmath}
\usepackage{natbib}
\usepackage{hyperref}
\usepackage{graphicx}
\usepackage{color}
\usepackage{amssymb}
\usepackage{exscale,relsize,supertabular}
\usepackage{ifpdf} 

\begin{document}
\title{Two Microspheres in an External Flow: a Dance of Cause and Effect}

\author{Golnaz Najafi Gol-Vandani}
\affiliation{Department of Physics, University of Guilan, P.O. Box 41335-1914, Rasht, Iran}
\affiliation{co-first authors}

\author{Simone Di Leo}
\affiliation{Department of Physics, Cavendish Laboratory, University of Cambridge, Cambridge CB3 0HE, United Kingdom}
\affiliation{co-first authors}

\author{Jurij Kotar}
\affiliation{Department of Physics, Cavendish Laboratory, University of Cambridge, Cambridge CB3 0HE, United Kingdom}

\author{Pietro Cicuta}
\email{pc245@cam.ac.uk}
\affiliation{Department of Physics, Cavendish Laboratory, University of Cambridge, Cambridge CB3 0HE, United Kingdom}

\author{Seyyed Nader Rasuli}
\email{rasuli@ipm.ir}
\affiliation{Department of Physics, University of Guilan, P.O. Box 41335-1914, Rasht, Iran}
\affiliation{School of Physics, Institute for Research in Fundamental Sciences (IPM), P.O.Box 19395-5531, Tehran, Iran}

\begin{abstract}
In low Reynolds number swimming and pumping,  differently to everyday experience,  a net motion (or flow)  can  be achieved only if the constructing parts of the swimmer (or pump)  follow a non-trivial pattern of motion, in order to break {\em time reciprocity}.  The case of a driven  fan, which spins to create a flow of air, but conversely rotates when turned off and subjected to a strong external flow,  is a  familiar example of reciprocal connection between {\em physical cause} and {\em effect}.   We explore here in a well controlled low Reynolds number system  whether such an exchange of the  {\em cause} and {\em effect} also holds in the low Reynolds number regime. As a case study we investigate  the motion of two microspheres which interact hydrodynamically through their surrounding fluid. Each sphere is constrained in a fixed optical trap potential, allowing local fluctuations around an equilibrium position. An {\em external flow} is shown to induce non-trivial coupled motion. We find a signature of reciprocity:  the nonequilibrium sphere fluctuations  mimic the symmetry of the motions that one would impose in order for them to produce a constant flow.
\end{abstract}


\maketitle

\begin{figure}[t!]
\includegraphics[width=1.0\columnwidth]{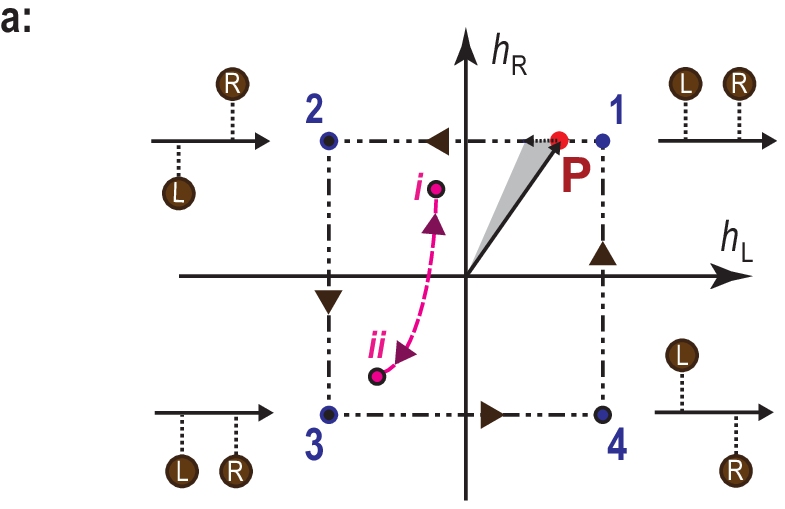}\\
\includegraphics[width=1.0\columnwidth]{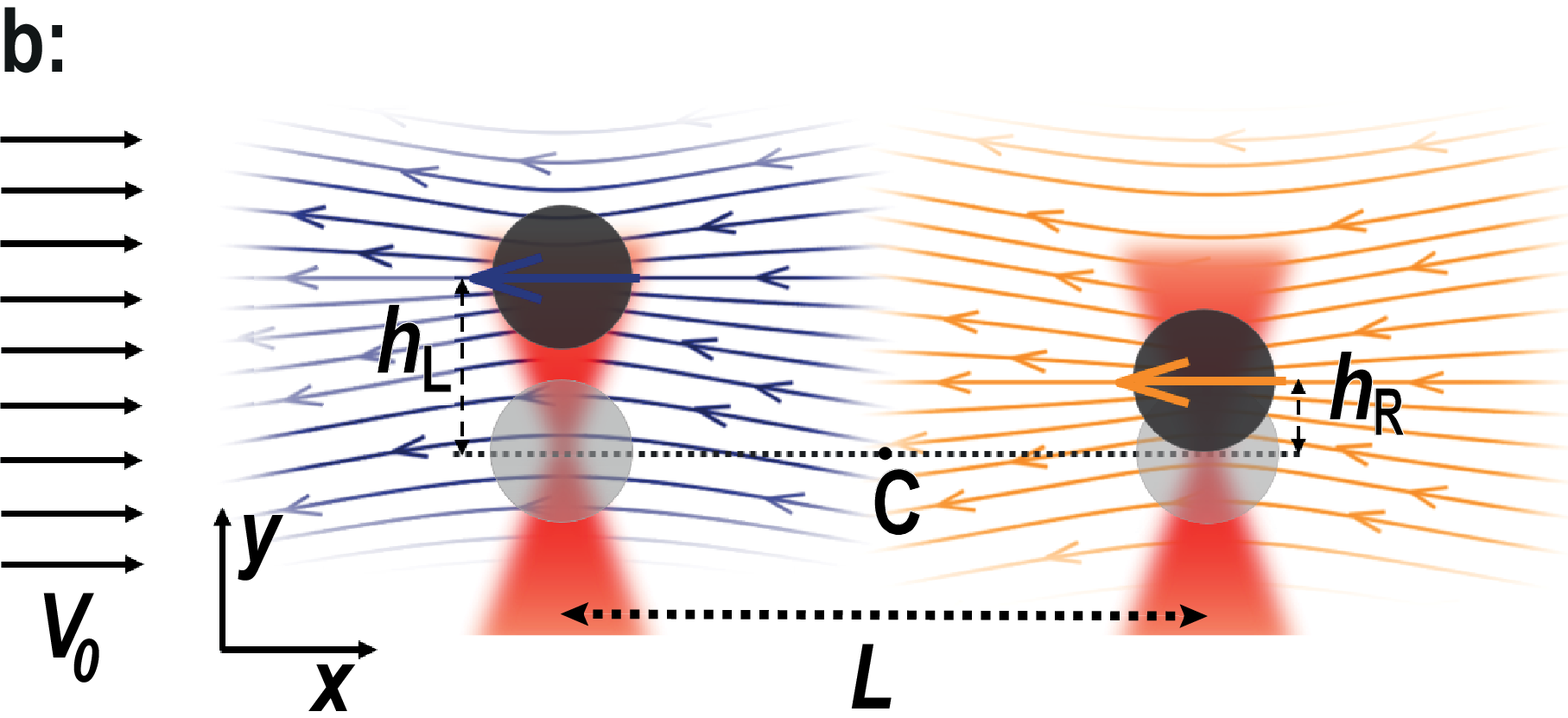}
\caption{A  two sphere system for investigating non-equilibrium fluctuations. ({\bf a})~2D {\em shape-space} of a pair of spheres which displace vertically. Two closed paths are marked as examples: A `go and return' path from $i$ to $ii$, is a reciprocal configuration change; its full cycle produces no net flow. In contrast the cycle of $1\rightarrow 2\rightarrow 3\rightarrow 4\rightarrow 1$  is non-reciprocal, it connects four states of the Left and Right spheres, as illustrated at the four corners of the path. The state point $\bf P$ sweeps the enclosed area by a rate of $\protect\overrightarrow{\,(h_{\mathsf L}, h_{\mathsf R})\,}\times \protect\overrightarrow{(\partial_t h_{\mathsf L}, \partial_t h_{\mathsf R})}/2 = (h_{\mathsf L}\,\partial_t h_{\mathsf R} - h_{\mathsf R}\, \partial_t h_{\mathsf L})/2$. The produced flow is rightward, and proportional to the average sweeping rate. ({\bf b})~Schematic of the experiment: two micron-size spheres (black disks) are optically trapped. Their vertical distances from the traps centers, $h_{\mathsf R}$ and $h_{\mathsf L}$, fluctuate due to the thermal noises, and restore to the equilibrium positions (gray disks) due to the optical trap harmonic force. An applied horizontal flow modifies the hydrodynamic interaction between the two spheres, and consequently modifies their height fluctuations.  We investigate if this modification can break time reciprocity, showing that the spheres follow trajectories similar to those required to produce a net flow in a quiescent fluid.}
\label{Fig-suggested-pump}
\end{figure}

Swimming and pumping at  and below the micron scale take place at low Reynolds number  (Re), and are typical in many technological and  biological systems~\cite{Lauga09,bray00}.  In contrast to the behaviour of larger scale systems, where inertia usually dominates, achieving a net motion depends here on the temporal order of the {\em displacements of swimmer's components}, and not on how fast they have been displaced. This is due to the linearity of flow, well described by Stokes' creep flow equations, ensuring for example that moving one component {\em fast} and  returning it back {\em slowly} produces a zero net momentum transfer into the fluid~\cite{Purcell77,Wilczek87,Wilczek89}. Consequently, a low Re swimmer/pump  needs more than one degree of freedom to function~\cite{Purcell77,Wilczek87,Wilczek89,cicuta10b}. This has lead the community to propose a family  of minimalistic {\em low Re} swimmers/pumps with two degrees of freedom~\cite{Stone03,Golestanian04,cicuta09a,cicuta10b}; they follow simple but non-trivial patterns of shape changes, breaking time reversal symmetry, thus causing a swim/pump action.  The configuration state of these minimal systems can be characterized by a point in a 2D {\em shape space}~\cite{Wilczek87,Wilczek89}, where a {\em reciprocal shape change} means that the point follows some path, returning back on along the same path; such {\em go and return} yields no net motion.  A {\em non-reciprocal} shape change corresponds instead to a path enclosing a finite area in the shape space, as shown in Fig.~\ref{Fig-suggested-pump}(a). It has been shown in many cases that the net motion/flow of a swimmer/pump is linearly proportional to that enclosed area~\cite{golestanian2008analytic,alouges2008optimal}.

In this Letter we ask: if we were able to put one of the minimal swimmers/pumps in an external flow, would its components follow the very same non-reciprocal pattern of motion that they would need to have when swimming/pumping on their own? Would this show up in correlations of the weakly non-equilibrium fluctuations within the system?
To quantify what we think of as a {\em reverse dance in an external flow}, we study a very simple two spheres system (see Fig.~\ref{Fig-suggested-pump}(b)): Two spheres of radius $a$ are held in optical traps at a fixed horizontal separation of $L$. We monitor the {\em vertical} displacements $h_{\mathsf R}$, $h_{\mathsf L}$ of the right and left spheres from their average positions (with the fluid at rest, the average positions are the trap minima). The state of the system is expressed by the point $P$ in the shape space (see Fig.~\ref{Fig-suggested-pump}(a)). If we consider the cycle through points (1,2,3,4,1), the rate by which $P$ sweeps the enclosed area,  averaged temporally over a full cycle, is:
\begin{eqnarray}\label{Eq:S}
S = \langle\, h_{\mathsf L}\,\partial_t h_{\mathsf R} - h_{\mathsf R}\, \partial_t h_{\mathsf L}\,\rangle/2.
\end{eqnarray}
$S$ is the key to estimate the {\em average induced velocity field}, due to the sequence of spheres' displacements in that cycle, as can be seen by obtaining $\langle\vec{V}_{\rm C}\rangle$, the average induced velocity at the point $\rm C$ in the middle of two spheres (see Fig.~\ref{Fig-suggested-pump}(b)). The hydrodynamic force induced by either of the spheres is proportional to its radius; likewise  the induced fluid velocity due to the sphere's motion  depends linearly on $a$. A single sphere, however, cannot produce a net flow by its cyclic motion; there should be another displacing sphere which resists and modifies its induced velocity field, as in~\cite{cicuta10b}. A simple scaling argument clarifies the factors in play:  $\langle\vec{V}_{\rm C}\rangle$ would scale like $a \times a$, therefore $\langle\vec{V}_{\rm C}\rangle \propto a^2 S$. What remains is the correct dimension,  we need a length cube in the denominator to fulfill the correct dimension of $\text{length}/\text{time}$; the only relevant length scale, at the point $\rm C$, is $L$. So $\langle\vec{V}_{\rm C}\rangle\propto a^2 S/L^3$. A detailed calculation is carried out in Supplementary Materials and gives the prefactor: $\langle\vec{V}_{\rm C}\rangle =9a^2\,S/L^3$. This simple example  shows how $S \neq 0$ is the signature of a non-reciprocal displacement. Accordingly, we refine the question of the reverse dance of two spheres: Given two spheres  trapped about $h_{\mathsf R}=h_{\mathsf L}=0$, does an external fluid flow modify their vertical fluctuations, so that a non-zero value for $S$ results?

\paragraph{Bead dynamics.} Without any loos of generality, see~\cite{Sup_Mat}, the Langevin equation for bead dynamics~\cite{edwards86}  can be written, in the {\em low Reynolds number} regime, for the vertical direction {{\em i.e.}\,$\hat{y}$} as:
\begin{eqnarray}\label{Eq:Force-Bal-Z}
0 = - K_{y}\,h_{i} + F_{i,y} + \zeta_{i,y}(t).
\end{eqnarray}
This force equation  balances on the $i$th bead the externally applied forces,  with $K_{y}$ the trap stiffness in the $y$ direction (and $i\in\{{\mathsf L},{\mathsf R}\}$ is the bead's index), with the hydrodynamic forces $F_{i,y}$ and Brownian forces $\zeta_{i,y}(t)$ ($y$ components). Note that in the presence of more than one bead, both $F$ and $\zeta$ are functions of all bead positions and velocities: As $\text{Re}\ll 1$, the fluid dynamics is governed by the linear Stokes equation~\cite{brenner83}, thus the fluid velocity
at each particle is the linear combination of the external flow ({\em i.e.}$\,V_{0}\,\hat{x}$), added to modifications caused by the {\em hydrodynamic forces} that either of the beads exerts to the fluid, $-\vec{F}_{j}$. This is formalized as:
\begin{eqnarray}\label{Eq:Flow-onm-spheres}
\frac{\partial}{\partial t}\vec{r}_{i}= V_{0}\,\hat{x} + \sum_{j=\{{\mathsf R},{\rm L}\}} {\bf{H}}_{i,j}\times(- \vec{F}_{j}).
\end{eqnarray}
The $\vec{r}_{i}=x_{i}\,\hat{x}+y_{i}\,\hat{y}+h_{i}\hat{z}$ measures the $i$th bead displacement from its equilibrium point, and ${\bf{H}}_{i,j}$ is the Oseen
tensor~\cite{oseen1927neuere,brenner83,russell}. If the beads are much smaller than their separation ({\em i.e.} $a\ll L\,$), then ${\bf{H}}_{i,j}$ has the form:
\begin{eqnarray}\label{Eq:Oseen's-Tensor}
\!\!\bf{H}_{i,j}=
\begin{cases}
\rule{0pt}{4ex}\mathlarger{\frac{\textbf{I}}{6 \pi \eta {\it{a}}}} \qquad\qquad\qquad\qquad\quad i=j\\
\rule{0pt}{4ex}\mathlarger{\frac{1}{8 \pi \eta }\times\frac{\textbf{I}+\hat{e}_{ij} \hat{e}_{ij}}{r_{ij}}} \quad\quad\qquad i\neq j \,,
\end{cases}
\end{eqnarray}
where $\textbf{I}$ is the $3\times3$ unity matrix, $\eta$ the fluid's viscosity, and $\vec{r}_{ij}=r_{ij} \hat{e}_{ij}$ the vector which connects the i$th$  to the $j$th sphere, $r_{ij}=|\vec{r}_{ij}|$, and $\hat{e}_{ij}=\vec{r}_{ij}/r_{ij}$. We can use Eq.(\ref{Eq:Flow-onm-spheres}) and Eq.(\ref{Eq:Oseen's-Tensor}) to solve the hydrodynamic forces~\cite{Sup_Mat} and make explicit the coupled Langevin equations. For spheres widely apart (compared to their radius $a \ll L$, and vertical fluctuations $h_{i} \ll L$) we obtain:
\begin{eqnarray}\label{Eq:dynamics-of-h}
\partial_{t}h_{i} = \mathlarger{\epsilon}\, \partial_{t}h_{\rm j} + \mathlarger{\epsilon}(\frac{h_{\mathsf L}-h_{\mathsf R}}{L})\,V_{\circ} - \frac{h_{i}}{\tau} + \frac{\zeta_{\rm i, y}(t)}{{6\pi\eta\,a}},
\end{eqnarray}
where $i$ indexes either of two beads, and $j \neq i$ the other one.
Here $\tau = 6\pi\eta\,a/K_{y}$ is a bead's relaxation time inside an isolated trap, and ${\mathlarger{\epsilon}}=3a/4L$ is a dimensionless coupling parameter which quantifies how the vertical motion of one bead induces
motion in the surrounding fluid, hence applying a force to the other bead.
Finally, the Brownian forces have zero mean $\langle\zeta_{\rm i}(t)\rangle=0$ and satisfy the fluctuation-dissipation
theorem $\langle\zeta_{\rm i}(t)\,\zeta_{\rm j}(t')\rangle=2k_{\rm B}T \,\bf{H}^{-1}_{i,j}\delta(t-t')$~\cite{edwards86,russell}, where $T$ is the temperature of the fluid and $k_{\rm B}$  is Boltzmann's constant.

Various averages, autocorrelations and crosscorrelations can be evaluated within this new condition of trapped beads subject to flow.  Firstly,  taking the average of Eq.(\ref{Eq:dynamics-of-h}), one  obtains $\langle h_{\mathsf L}\rangle=\langle h_{\mathsf R}\rangle=0$~\cite{Sup_Mat}, i.e.  neither the external flow nor the hydrodynamic interactions among the beads cause a shift in the vertical position.
Secondly,  we {\em analytically} solve Eq.(\ref{Eq:dynamics-of-h}) for both beads~\cite{Sup_Mat} to obtain $2$nd order correlations. For example, the left bead's auto-correlation $\langle h_{\mathsf L}(t + \Delta t)h_{\mathsf L}(t)\rangle$, is:
\begin{eqnarray}\label{Eq=auto-cor-L}
\hspace{-0.45cm}\!&&\!\!\!\!\!\frac{k_{\rm B} T}{K_y} {\mathlarger{\biggr[}}
\frac{1}{2}\mathlarger{(e^{-\Delta t\,/\tau(1-\epsilon)} + e^{-\Delta t\,/\tau(1+\epsilon)})}\nonumber\\
\hspace{-0.45cm}\!&&\!\qquad +\frac{V_{\circ}\tau}{2\,L}\mathlarger{(}(1+2\epsilon) \mathlarger{e^{-\Delta t\,/\tau (1+\epsilon)}\!-\! e^{-\Delta t\,/\tau(1-\epsilon)})} {\mathlarger{\biggr]}}.
\end{eqnarray}
The first terms are the well-known result of Meiners and Quake~\cite{quake99}, showing how hydrodynamic interactions affect auto-correlations in the thermal fluctuations of trapped beads, in a quiescent fluid. The result is here modified by the external fluid flow, with an additive term proportional to $V_{\circ}$. We obtain $\langle h_{\mathsf R}(t + \Delta t)h_{\mathsf R}(t)\rangle$ the same as $\langle h_{\mathsf L}(t+ \Delta t)h_{\mathsf L}(t)\rangle$, with a mere change of $V_{\circ} \rightarrow -V_{\circ}$~\cite{Sup_Mat}, consistent with the fact that the only source of symmetry breaking is the external fluid flow.   An observer sitting in the reference frame of the moving fluid would distinguish the beads as the {\em front bead}, and the {\em back bead}, see Fig.~\ref{Fig-suggested-pump}. Reversing the fluid flow simply switches the front and back beads, hence the physics they obey; consequently the change of $V_{\circ} \rightarrow -V_{\circ}$ simply switches the expression for the right and left beads.

Eq.(\ref{Eq=auto-cor-L}) yields $\langle h_{\mathsf L}^2(t)\rangle =(k_{\rm B}\,T/K_y)\times(1+\epsilon V_{\circ}\tau/L)$: this means that vertical fluctuations of the left sphere, {\em i.e.} $\langle h_{\mathsf L}^2\rangle - \langle h_{\mathsf L}\rangle^2$, are modified by a factor of $(1+\epsilon V_{\circ}\tau/L)$, and similarly $\langle h_{\mathsf R}^2\rangle - \langle h_{\mathsf R}\rangle^2$ is corrected by $(1-\epsilon V_{\circ}\tau/L)$. So a left to right flow ({\em i.e.}$\,V_{\circ}>0$) amplifies the left bead's fluctuations, and reduces those of the right one.  The external flow and the hydrodynamic interactions can be interpreted as causing an effective softening of the upstream trap, while the downstream trap is strengthened. To understand the underlying physics, we approximate each bead to a {\em point force}, sitting {\em on average} at its equilibrium position of $h=0$, and resisting fluid's flow of $V_{\circ}\hat{x}$ with a drag force of $\vec{f}=-6\pi\eta\,a V_{\circ}\hat{x}$. The point force induces a diverging pattern of flow in its front, and a converging one in its back. The left bead then experiences the diverging fluid flow produced by the right one (see orange lines in Fig.~\ref{Fig-suggested-pump}(b)), the diverging flow tends to push it away from $h_{\mathsf L}=0$ and amplifies its fluctuations. Similarly, the right bead experiences a converging flow which diminishes its fluctuations.

\paragraph{Reverse dance.} The most striking symmetry breaking, and the main result of our work, comes from calculating the {\em cross correlations} of the fluctuating beads (full algebra in~\cite{Sup_Mat}):
\begin{eqnarray}
\hspace{-6.15cm}&&\hspace{-.8cm} \langle h_{\mathsf L}(t+\Delta t)h_{\mathsf R}(t) \rangle = -\frac{k_{\rm B} T}{2K_y}
{\mathlarger{\biggl[}}\mathlarger{(1+\frac{V_{\circ}\tau}{L})}\times \nonumber\\
&& \hspace{+1.10cm}\mathlarger{(e^{-\Delta t\,/\tau(1+\epsilon)}\! -\! e^{-\Delta t\,/\tau(1-\epsilon)})}{\mathlarger{\biggr]}}\!+\!O(\epsilon^2).
\label{eq:cross_corr_fluid_lr2}
\end{eqnarray}
This extends the Meiners and Quake result~\cite{quake99}, accounting for the external flow, resulting in the pre-factor term $(1+V_{\circ}\tau/L)$. It is  $(1-V_{\circ}\tau/L)$ for the other cross correlation, $\langle h_{\mathsf R}(t+\Delta t)h_{\mathsf L}(t) \rangle$~\cite{Sup_Mat}, again a consequence of the fact that exchanging left and right beads is equivalent to a change of $V_{\circ}\rightarrow-V_{\circ}$. These expressions give us a  quantity that we can also measure experimentally,  {\em measured $S$}:
\begin{eqnarray}
\hspace{-.17cm}S_{\rm meas}\! = \!\lim_{\Delta t \rightarrow 0} \!\frac{\langle h_{\mathsf R}(t+\Delta t)h_{\mathsf L}(t) \rangle - \langle h_{\mathsf L}(t+\Delta t)h_{\mathsf R}(t)\rangle}{2\Delta t}.
\label{Eq:diff2CC}
\end{eqnarray}
Using $\lim_{\Delta t \rightarrow 0}\mathlarger{(}e^{-\Delta t\,/\tau(1+\epsilon)}-e^{-\Delta t\,/\tau(1-\epsilon)}\mathlarger{)}/2\Delta t=\epsilon/\tau$ we have the theoretical expectation from the Langevin calculus:
\begin{eqnarray}\label{Eq:S_analytic}
S_{\rm meas} =\frac{k_{\rm B} T}{K_y} \times \frac{\epsilon V_{\circ}}{L} = \frac{3}{4}a V_{\circ} \times \frac{\langle h^2 \rangle_{0}}{L^2}.
\end{eqnarray}
\begin{figure}
\includegraphics[width=1.0\columnwidth]{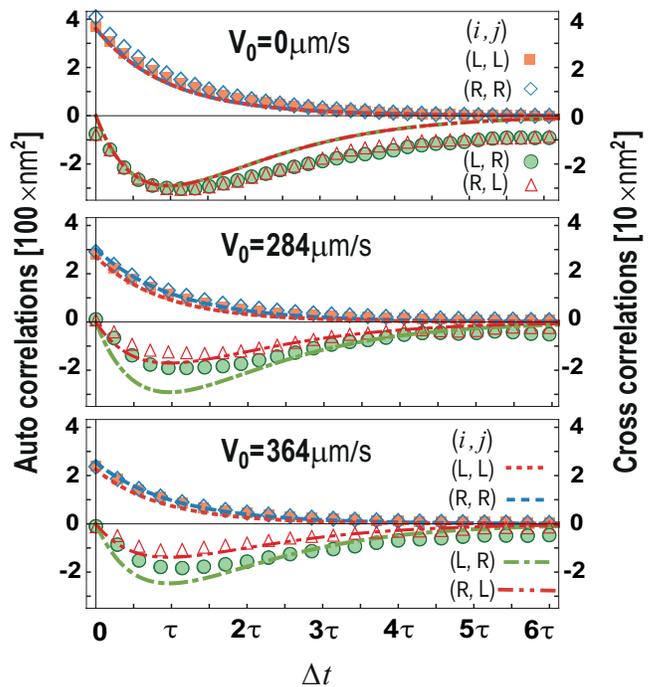}
\caption{The presence of an external flow breaks the left-right symmetry in the fluctuations of trapped beads.  Auto- and cross-correlations {\em i.e.} $\langle h_{i}(t+\Delta t)h_{j}(t) \rangle$, versus $\Delta t$, are plotted for various values of external drive, $V_{0}$. Beads' separation is fixed to $L=6\,\mu$m. Two cross correlations clearly separate from each other, as $V_{0}$ increases; the same happens for two auto-correlations, but in a smaller scale! The reason is that the external drive, which breaks right-left symmetry, contributes in the cross correlations like $V_{\circ}\tau/L$, {\em c.f.} Eq.(\ref{eq:cross_corr_fluid_lr2}), whereas it contributes in the auto correlations with a smaller factor of $\epsilon V_{\circ}\tau/L$.
The time lags are expressed in the natural relaxation time of each experiment; the trapping stiffness is $K_y=11.3, 14.2$ and $17.0$\,pN$/\mu$m from top to bottom. }\label{Fig:Auto_Cross_Cor}
\end{figure}
Eq.(\ref{Eq:S_analytic}) shows that $S_{\rm meas}$ linearly depends on $V_{\circ}$. We find particularly interesting the fact that if we were intentionally displacing beads (in a  $V_{\circ}=0$ background), then  $S$ would quantify the induced fluid motion (the pumping effect). Reversing the cycle ($S\rightarrow -S$) would simply reverse the induced flow. Now, if in this regime the `cause' and `effect' can be reversed, we would expect that reversing $V_{\circ}$ would result in reversing $S_{\rm meas}$. It means that $S$ can not depend on $V^2$ (or any other even power), as indeed in Eq.(\ref{Eq:S_analytic}).
It is also useful to point out that  $S_{\rm meas}$ depends on $\langle h^2 \rangle_{0}=k_{\rm B} T/K_y$, i.e. on the original amplitude of {\em thermal fluctuations}, thus  $S_{\rm meas}=0$ if the beads were rigidly fixed, or lacking any intrinsic fluctuations.

These considerations have been put to experimental test: Building on our previous methods~\cite{cicuta09a,bruot13r}, the displacement of the particles is captured at 400 to 1000 frames per second (depending on  the dimension of the recorded region). Videos were recorded for 6~minutes, corresponding to  between 144,000 and 360,000 frames. The laser trap is well approximated by a harmonic potential, as can be seen by  the distribution of particle positions; this also allows calibration of the trap stiffness, as detailed in~\cite{Sup_Mat}.
The trapping laser is time-shared by steering through a pair of acoustic-optical deflectors (AOD). Spherical silica  colloidal beads from Bangs Laboratories, with density  2\,g/cm$^3$, refractive index $\approx$1.45 and diameter of $2a=3.47$\,$\mu$m, are
dispersed in a solution of water/glycerol with a viscosity $\eta =2.4$\,mPas in the flow experiments. In contrast to much work in the field (including all our previous work) this experiment requires an open channel, which  introduces spurious velocity fluctuations from ambient pressure variation. In~\cite{Sup_Mat} we describe the characterization of this, and also of the velocity profile induced by the syringe pump (Legato 110, KD scientific, USA).

\begin{figure}
\includegraphics[width=0.99\columnwidth]{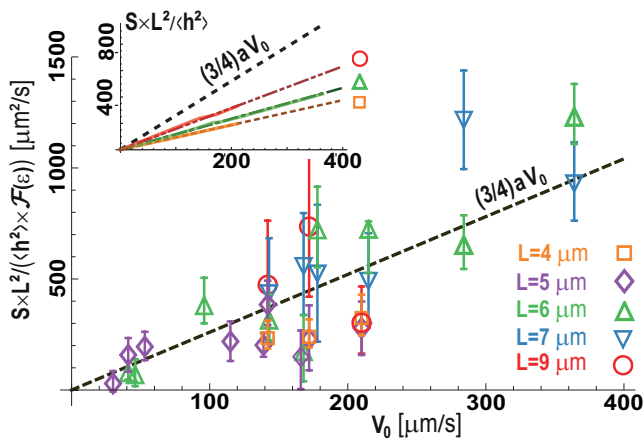}
    \caption{Experimental data for a variety of applied flow velocities $V_\circ$ and inter-bead separations  collapses to the  theoretical master curve: The  sweeping parameter $S$ is    scaled by $\langle h^2 \rangle_{0}/L^2$ and by the hydrodynamic correction ${\mathcal F}(\epsilon)$.   The  collapse shows  a linear dependence on $V_\circ$  of all the experimentally measured  $S$, obtained with various stiffness of the trap and initial separation, $L$, consistent with the theory developed in this letter.
    The inset discloses the effect of near-field corrections: Without the ${\mathcal F}(\epsilon)$ scaling, analytics and simulation clearly deviate from the far field prediction, the dashed line. The {dotted} lines are analytical results, verified by thick data from our full stochastic simulation, for three values of $L$ as labelled.
}
\label{Fig:Sy*ky*L^2}
\end{figure}

\paragraph{The dance of real beads.}
So far, we have modeled the beads' hydrodynamics interaction using the far field approximation, {\em i.e.} each bead is a point force inside fluid. However, in our experimental setup, the beads' horizontal separations were even reduced to about $L \simeq 4\mu$m, barely larger than $2a$.
This forces us to consider the near field hydrodynamics of the beads. We thus refine the hydrodynamics: starting from the known correction for hydrodynamic interaction of two spheres up to $\epsilon^6 = (3a/4L)^6$, which holds in special cases~\cite{brenner83}, we extend it to arbitrary situation~\cite{Sup_Mat}. This corrects Eq.(\ref{Eq:S_analytic}) to:
\begin{eqnarray}
S =  \frac{3}{4}a V_{\circ} \times{\mathcal F}(\epsilon) \frac{\langle h^2 \rangle_{0}}{L^2},
\label{Eq:Accurate_S}
\end{eqnarray}
where ${\mathcal F}(\epsilon)=1 -3 \epsilon + (22/9) \epsilon ^2 - (260/27)\epsilon ^3 +(904/27) \epsilon ^4 + (725/27) \epsilon ^5$ is obtained using higher order corrections to the Oseen tensor~\cite{brenner83,Sup_Mat}. ${\mathcal F}(\epsilon)$ is clearly $=1$ for far separated beads ({\em i.e.} $\epsilon \ll 1$), but deceases to $\sim 0.4$ as two beads approach each other. Fig.~\ref{Fig:Sy*ky*L^2} shows the scaled sweeping parameter, $S\times L^2/({\mathcal F}(\epsilon)\,\langle h^2 \rangle_{0})$, versus the external induced velocity. The dashed (black) line depicts the refined analytic prediction of Eq.(\ref{Eq:Accurate_S}); it sits well between the experimental data.

With this complex hydrodynamics, we also carried out a {\em stochastic simulations}~\cite{ermak1978,gillespie1996mathematics}. The inset of Fig.~\ref{Fig:Sy*ky*L^2} compares the analytic results of Eq.(\ref{Eq:Accurate_S}) with their corresponding simulation results, for selected values of $L$. Interestingly, the full stochastic simulation confirms the linear dependence of $S$ on the fluid velocity, verifying the aforementioned argument on the its linear dependence.

\paragraph{A dance to reduce the resistance}
The results so far illustrate a clean example of reciprocity in cause and effect, between the conformations necessary to induce a flow (with active forces, thus dissipating energy), and the correlations in the conformation fluctuations observed when the same flow is applied externally (a non-equilibrium situation, again with dissipation). There is but one final question to establish a consistent picture: Are these correlated fluctuations a dance in favor of the driven flow, or against it? Two unmoving beads, resisting against the fluid flow, experience a total drag of $\langle F_{\rm drag} \rangle =2\times 6\pi\eta a V_{\circ}\,( 1 - 2\epsilon )$ by the fluid, which results in a corresponding energy dissipation. When the beads can fluctuate, their vertical dance corrects $\langle F_{\rm drag} \rangle $ to \cite{Sup_Mat}:
\begin{equation}
\!\langle F_{\rm drag}\rangle =2\times 6\pi\eta a V_{\circ}\mathlarger{(} 1 - 2\epsilon \mathlarger{)} - 6\pi\eta a \times \frac{3 a}{2 L^2} S +O(\epsilon^3).
\end{equation}
As $S>0$, there is a reduction  in the drag, and correspondingly in the dissipation. Using Eq.(\ref{Eq:S_analytic}) we find that $F_{\rm drag}$ has been reduced by a factor of $(1 - \epsilon^2 \langle h^2 \rangle_{0}/L^2)$. For the system we studied $\langle h^2 \rangle_{0}/L^2 \approx 10^{-4}$, {\em c.f.} Fig.~\ref{Fig:Auto_Cross_Cor} for $\langle h^2 \rangle$. However, if we scale down this system to nano sizes, where $k_{\rm B}T$ induces comparatively much  more violent fluctuations, the term $\langle h^2 \rangle_{0}/L^2$ becomes significant. Moreover, as the beads are really allowed to dance in all three dimensions ({\em i.e.} also $x$ and $z$), the reduction factor of $\epsilon^2 \langle h^2 \rangle_{0}/L^2$, also amplifies by a factor of $6$~\cite{Sup_Mat}.

 This experiment (and theory) interestingly suggests that the intrinsic thermal fluctuations, when coupled with hydrodynamic interactions, cause  a dance in favour of driven flow; and whenever such fluctuations are comparable to other length scales in the problem, that dance would {\em non-negligibly} reduce the rate of heat production. Moreover, it suggests that the complicated swimming strategies at low Reynolds number found during the last two decades (e.g. cyclic motions of three sphere systems), were probably already  spontaneously happening in nature.
One open question is how wide is the class of  non-equilibrium systems that share the physics illustrated in this two bead system.

\begin{acknowledgements}
    We are grateful to Marco Cosentino Lagomarsino, Evelyn Hamilton and Luigi Feriani for very useful discussions, and Ramin Golestanian for his constructive criticism. PC and JK acknowledge funding from ERC CoG HydroSync, and SDL an EU Erasmus studentship.
\end{acknowledgements}


\end{document}